\documentclass[twocolumn,aps,prl,superscriptaddress,showpacs]{revtex4-1}
\usepackage[dvips]{graphicx}
\usepackage{amssymb,amsfonts,amsmath}
\usepackage{epsfig}
\scrollmode

\begin{document}

\title{On ergodic least-squares estimators of the generalized diffusion 
coefficient for fractional Brownian motion}

\author{Denis Boyer}
\email{boyer@fisica.unam.mx}
\affiliation{Instituto de F\'{\i}sica, Universidad Nacional Aut\'onoma de M\'exico,
D.F. 04510, Mexico}
\author{David S. Dean}
\email{david.dean@u-bordeaux1.fr}
\affiliation{Universit\'e de  Bordeaux and CNRS, Laboratoire Ondes et
Mati\`ere d'Aquitaine (LOMA), UMR 5798, F-33400 Talence, France}
\author{Carlos Mej\'{\i}a-Monasterio}
\email{carlos.mejia@upm.es}
\affiliation{Laboratory of Physical Properties, Technical University
of Madrid, Av. Complutense s/n 28040, Madrid, Spain}
\affiliation{Department of Mathematics and Statistics, University of Helsinki,
 P.O.  Box 68 FIN-00014, Helsinki, Finland}
\author{Gleb Oshanin}
\email{oshanin@lptmc.jussieu.fr}
\affiliation{Laboratoire de Physique Th\'eorique de la Mati\`ere
Condens\'ee (UMR CNRS 7600), Universit\'e Pierre et Marie Curie, 4
place Jussieu, 75252 Paris Cedex 5 France}

\date{\today}

\begin{abstract}
  We  analyse  a class  of  estimators  of the  generalized  diffusion
  coefficient  for fractional  Brownian  motion $B_t$  of known  Hurst
  index $H$, based  on weighted functionals of the  single time square
  displacement.   We show  that for  a  certain choice  of the  weight
  function  these functionals  possess  an ergodic  property and  thus
  provide   the   true,   ensemble-averaged,   generalized   diffusion
  coefficient  to any  necessary  precision from  a single  trajectory
  data,  but  at  expense   of  a  progressively  higher  experimental
  resolution. Convergence is fastest  around $H\simeq0.30$, a value in
  the subdiffusive regime.
\end{abstract}


\maketitle

Single molecule  spectroscopy techniques allow the  tracking of single
particles over  a wide  range of time  scales \cite{bra,saxton,mason}.
In complex media such as living cells, a number of recent studies have
reported  evidence  for   subdiffusive  transport  of  particles  like
proteins   \cite{weber},   viruses   \cite{seisenberger},   chromosome
monomers  \cite{bronstein},  mRNA  \cite{golding}  or  lipid  granules
\cite{ralf}.  Subdiffusion  is typically characterized  by a sublinear
growth   with   time   of   the  mean   square   displacement   (MSD),
$\mathbb{E}(\mathbf{B}_t^2)=K   t^{\nu}$  with   $\nu<1$,  where
$\mathbf{B}_t$  is the  particle  position at  time $t$,  $\mathbb{E}$
denotes the ensemble average and $K$ is a generalized diffusivity.

A  growing  body  of  single  trajectory  studies  suggest that  fractional
Brownian motion (fBm), among  the variety of stochastic processes that
produce subdiffusion, may be a  model particularly relevant to subcellular
transport.   FBm is  a  Gaussian continuous-time  random process  with
stationary increments and is  characterized by a so-called Hurst index
$H=\nu/2$. If  $H<1/2$, trajectories  are subdiffusive with  increments that
are  negatively  and long  range  correlated \cite{mandelbrot}.   Such
correlations   were   observed    in   subdiffusing   mRNA   molecules
\cite{polaco},  RNA-proteins or  chromosomal loci  \cite{weber} within
{\it E.  coli} cells.  Similarly,  fBm can be used to describe the dispersion of
apoferritin proteins in  crowded dextran solutions \cite{weissprl} and
of lipid molecules in lipid bilayers \cite{metzlerlipids}.

Whereas the determination of an anomalous exponent from 
data has been extensively studied, as it  demonstrates deviation from standard Brownian motion (BM), 
the problem of  estimating the generalized diffusion constant $K$ has  received much
less attention. It appears that  $K$ is much  more sensitive than $\nu$ to many 
biological factors and its precise determination can potentially yield valuable information 
about the kinetics of transcription, translation and other 
physico-biological processes. The generalized diffusivity of RNA molecules in bacteria is greatly affected  (either positively or negatively) by perturbations, for instance  treatment with 
antibiotic drugs, which have however a negligible effect on 
$\nu$ \cite{weber}. Likewise, the coefficient $K$ of lipids in 
membranes is strongly reduced by small cholesterol concentrations, 
whereas $\nu$ remains unchanged \cite{metzlerlipids}. 
In the context of search problems, a particle following a subdiffusive fBm 
actually explores the $3d$ space more 
compactly than a BM and can have a higher probability of eventually 
encountering a nearby target \cite{weissbiophysj}. The larger the value of $K$, the
faster this local exploration.

In  this paper,  generalizing  our previous  results  for standard  BM
\cite{dean1}, we  present a method  to estimate the  ensemble averaged
diffusivity $K$ from the analysis of single fBm trajectories of {\it a
  priori}  known anomalous  exponent.  Estimating  diffusion constants
from data is  not an easy task when trajectories  are few and ensemble
averages cannot  be performed.  BM  and fBm are ergodic  processes and
time averages tend  to ensemble averages, but convergence  can be slow
\cite{dengbarkai}.   For  finite  trajectories of  finite  resolution,
variations by orders of magnitude have been observed for estimators of
the normal diffusion coefficient obtained from single particles moving
along DNA  \cite{austin}, in the  plasma membrane \cite{saxton}  or in
the cytoplasm  of mammalian cells \cite{goulian}.   Large fluctuations
are also manifest in subdiffusive cases \cite{weber,metzlerlipids}.

A broad dispersion in the measures of the diffusion coefficient raises
important questions about  optimal fitting methodologies. A reliable estimator
must  possess an  ergodic property,  so that  its most  probable value
should  converge to  the true  ensemble average  independently  of the
trajectory  considered   and  its   variance  should  vanish   as  the
observation time  increases.  Recently, much effort  has been invested
in  the analysis  of this  challenging problem  and  several different
estimators   have  been   analyzed,  based,   e.g.,  on   the  sliding
time-averaged square displacement \cite{greb1,greb3}, mean length of a
maximal  excursion \cite{tej},  the  maximum likelihood  approximation
\cite{berglund,michalet,mb,boyer,boyer1}    and    optimal    weighted
least-squares functionals \cite{dean1}.

Our aim here is to determine an ergodic least-square estimator for the
generalized diffusion coefficient when the underlying stochastic motion is given 
by a fBm. The estimators considered here are single time
quantities, unlike others based on fits of two-time quantities such as
the time averaged MSD.

Let us  consider a fractional  Brownian motion $B_t$ in  one dimension
with $B_0=0$ and zero expectation value  for all $t \in [0, T]$, where
$T$  is the  total observation  time. The  covariance function  of the
process is given by \cite{mandelbrot}:
\begin{eqnarray}\label{covariance}
{\rm Cov}\left(B_t,B_s\right) &=& \mathbb{E}\{\left(B_t - 
\mathbb{E}\{B_t\}\right)\left(B_s - \mathbb{E}\{B_s\}\right)\}  \nonumber\\
&=& \frac{K}{2} \left(t^{2 H} + s^{2 H} - |t - s|^{2 H}\right)\,,
\end{eqnarray}
where $D(=K/2$) is the generalized diffusion 
coefficient and the Hurst exponent $H\in(0,1)$.
The Hurst index describes the raggedness of the resulting motion, 
with a higher value leading to a smoother motion.
Standard Brownian motion is a particular case of the fBm corresponding 
to $H = 1/2$. As already mentioned, for $H < 1/2$ the increments of the 
process are negatively correlated so that the fBm is subdiffusive. On the other hand, for $H > 1/2$ the increments of  the process are positively correlated and superdiffusive behavior
is observed.

We  consider  a  single   trajectory  $B_t$,  that  is,  a  particular
realization of  an fBm process  with a known  $H$, and write  down the
following weighted least-squares functional:
\begin{equation}
\label{1}
F = \frac{1}{2} \int^T_0 dt \, W(t) \, \left(B_t^2 - K_f \, t^{2 H}\right)^2,
\end{equation}
where $W(t)$  is some weighting  function to be  determined afterwards
and $K_f$  is a  trial parameter.   We call $K_f$  an estimate  of the
generalized diffusion coefficient from the single trajectory $B_t$, if
it  minimizes  $F$.   Calculating  the partial derivative  $\partial
F/\partial K_f$, setting  it to zero and solving  the resulting equation
for $u = K_f/K$, we  find the following least-squares estimator of the
generalized diffusion coefficient $K$:
\begin{equation}
\label{estim}
u \equiv \frac{K_f}{K} = \frac{1}{K} \frac{\int^T_0 dt \, 
\omega(t) B_t^2}{\int^T_0 dt \, t^{2 H} \, \omega(t)}\,,
\end{equation}
where we have introduced the notation
\begin{equation}
\omega(t) = t^{2 H} \, W(t) \,.
\end{equation}
Note that the  estimator $u$  measures  the ratio  of  the observed  generalized
diffusion coefficient  for a single  given trajectory relative  to the
ensemble-averaged value.  Moreover,  $\mathbb{E}\{u \} \equiv 1$ holds
for  any  arbitrary  $\omega(t)$,   making  it possible  to  compare  the
effectiveness of  different choices of $\omega(t)$.   It is worthwhile
remarking  that $u$ is  given by  a single  time integration  (a local
functional)  and  thus  differs  from  other  estimates  used  in  the
literature   which  involve  two-time   integrals  (see   {\it  e.g.},
\cite{dengbarkai}).

Further  on, from a  straightforward calculation  the variance  of the
estimator $u$ is, for arbitrary weight function $\omega(t)$,
\begin{equation}
\label{vfbm}
{\rm Var}(u) = \frac{1}{K^2} \frac{\int^T_0 \int^T_0 dt \, ds\, 
\omega(t) \, \omega(s) \, 
{\rm Cov}\left(B_t^2, B_s^2\right)}{\left(\int^T_0 dt \, t^{2 H} \, 
\omega(t)\right)^2} \,,
\end{equation}
where ${\rm Cov}\left(B_t^2,  B_s^2\right)$ is the covariance function
of a squared fBm trajectory
\begin{eqnarray}
{\rm Cov}\left(B_t^2, B_s^2\right) &=& \mathbb{E}\{\left(B_t^2 - 
\mathbb{E}\{B_t^2\}\right)\left(B_s^2 - \mathbb{E}\{B_s^2\}\right)\} \,.
\end{eqnarray}
This function  can be calculated  exactly using Eq.~(\ref{covariance})
to give
\begin{eqnarray}
{\rm Cov}\left(B_t^2, B_s^2\right) &=&  2 \, {\rm Cov}^2\left(B_t, B_s\right) 
\nonumber\\
&=& \frac{K^2}{2} \left(t^{2 H} + s^{2 H} - |t - s|^{2 H}\right)^2 \,.
\end{eqnarray}
Inserting  the latter  expression into  Eq.~(\ref{vfbm})  and noticing
that the kernel is a symmetric function of $t$ and $s$, we have
\begin{equation}
\label{result}
{\rm Var}(u) = \frac{\int^T_0 \int^t_0 dt \, ds\, \omega(t) \, \omega(s) \, 
\left(t^{2 H} + s^{2 H} - (t - s)^{2 H}\right)^2}{\left(\int^T_0 dt \, 
t^{2 H} \, \omega(t)\right)^2} \,.
\end{equation}

Following Ref.\cite{dean1}, we choose
\begin{equation}
\omega(t) = (t_0 + t)^{-\alpha},\label{plw}
\end{equation} 
where  $t_0$ is  a lag  time and  $\alpha$ a  tunable exponent.   In a
discrete  time description,  $t_0$ can  be set  equal to  the interval
between successive measurements  \cite{dean1}.  We thus identify $t_0$
as  a  resolution parameter  in  the  present continuous  description. We also note that in  \cite{dean1},
it was proven that  a power law weight function of the type in Eq. (\ref{plw}) was optimal among all
weight functions. Fixing $t_0$ and  scanning over different values of  $\alpha$, we seek
the value  for which the variance  of $u$ is  smallest. Hopefully, for
such  value, the  variance should  vanishes in  the limit  of infinite
resolution  or infinite  data  size, {\it  i.e.}   when the  parameter
$\epsilon  = t_0/T$  tends  to zero.  To  check the  latter point,  we
consider  first the  limit  of an  infinitely  long observation  time,
$\epsilon =  0$.  For $\alpha  < \gamma_H= 1  + 2 H$ the  integrals in
Eq.~(\ref{result}) can be performed exactly yielding
\begin{eqnarray}
\label{var}
&&{\rm Var}(u) = \frac{\gamma_H - \alpha}{2} \Big(\frac{1}{1 - \alpha} 
+ \frac{2}{\gamma_H - \alpha} +\nonumber\\
&+& \frac{1}{2 \gamma_H -1  - \alpha} - 2 \frac{\Gamma(1 - \alpha) \, 
\Gamma(\gamma_H)}{\Gamma(1  + \gamma_H - \alpha)}\nonumber\\
 &+& \frac{\Gamma(1-\alpha) \Gamma(2 \gamma_H - 1) - 2 \Gamma(\gamma_H) 
 \Gamma(\gamma_H - \alpha)}{\Gamma(2 \gamma_H - \alpha)} \Big) \,,
\end{eqnarray}
where $\Gamma(\cdot)$  is the gamma-function.  On the  other hand, for
$\alpha  > \gamma_H  = 1  + 2  H$ and  $\epsilon =  0$, the  result in
Eq.~(\ref{result})  can  be   conveniently  represented  as  a  single
integral
\begin{eqnarray}
\label{var7}
&&{\rm Var}(u) = \frac{\Gamma\left(2 \gamma_H\right) \, 
\Gamma\left(2 \alpha - 2 \gamma_H\right) \, 
\Gamma^2\left(\alpha\right)}{\Gamma^2\left(\alpha - \gamma_H\right) \, 
\Gamma^2\left(\gamma_H\right)} \times \nonumber\\
&& \int^1_0 \left(1 + (1-x)^{2 H} - x^{2 H}\right)^2 \, _2F_1\left(\alpha, 2 
\gamma_H, 2 \alpha; x\right) \,,
\end{eqnarray}
where   $_2F_1\left(\cdot\right)$  is  the   confluent  hypergeometric
function.   The integral  in  Eq.~(\ref{var7}) can  be also  performed
exactly  by   using  the   series  representation  of   the  confluent
hypergeometric   function    and   then   resumming    the   resulting
series.  However, the  expression  obtained is  rather  lengthy as  it
contains           several           hypergeometric          functions
$_3F_2\left(\cdot\right)$. On  the other hand, the result  in the form
of  Eq.~(\ref{var7}) can be  tackled by  Mathematica; in  addition the
asymptotic behavior can be easily extracted from it, so that we prefer
to work with  the compact expression (\ref{var7}) rather  than with an
exact but cumbersome expression.

\begin{figure}[!t]
  \centerline{\includegraphics*[width=0.48\textwidth]{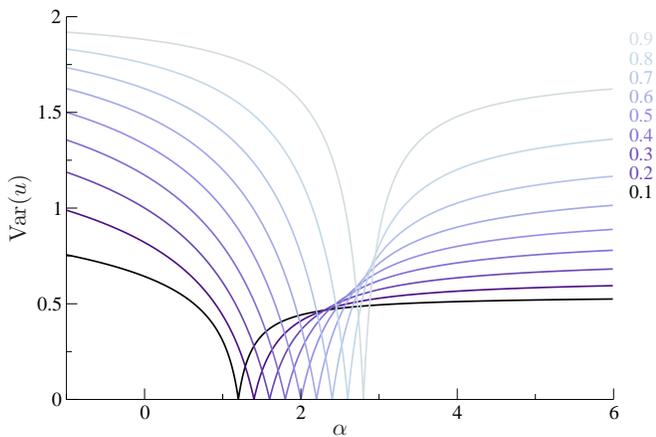}}
  \caption{(color  online)  The   variance  in  Eqs.~(\ref{var})  (for
    $\alpha < 1 + 2 H$) and (\ref{var7}) (for $\alpha > 1 + 2 H$) as a
    function of  $\alpha$, for different values of  the Hurst parameter
    $H$.}
\label{fig1}
\end{figure}

In  Fig.\ref{fig1} we  show  the  dependence of  the  variance of  the
estimator $u$  on the exponent  $\alpha$, for different values  of the
Hurst  index $H$.   We notice  that for  any fixed  $H$,  the variance
vanishes as $\alpha$ approaches $\alpha = 1 + 2 H$ and is non-zero for
any other value. This means that for a fractional Brownian motion with
Hurst  index $H$  the estimators  in Eq.~(\ref{estim})  with power-law
weight functions $\omega(t) =  (t_0 + t)^{-\alpha}$ possess an ergodic
property only when $\alpha = 1 + 2 H$.

\begin{figure}[!t]
  \centerline{\includegraphics*[width=0.45\textwidth]{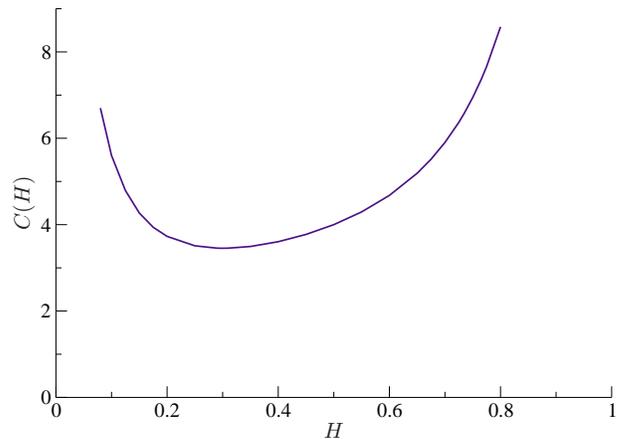}}
  \caption{Prefactor in Eq.~(\ref{finalvar}) as a function of the Hurst index.}
\label{fig2}
\end{figure}

The last  issue we discuss is that  of the decay rate  of the variance
when   $\epsilon$  is   small   but  finite   in   the  ergodic   case
$\alpha=1+2H$. It  is straightforward to  show from Eq.~(\ref{result})
that in  the limit $\epsilon \to  0$ the variance is  given to leading
order by:
\begin{equation}\label{finalvar}
{\rm Var}(u) \sim \frac{C(H)}{\ln(1/\epsilon)} \,,
\end{equation}
where $C(H)$ is a constant defined by:
\begin{equation}
C(H) = \int^1_0 \frac{dx}{x^{1 + 2 H}} \left(1 + x^{2 H} - 
(1-x)^{2 H}\right)^2 \,,
\end{equation}
which exists  for any $H \in  (0,1)$. This result generalizes  that of
Ref.  \cite{dean1} for ordinary Brownian  motion. We conclude that the
variance  of the  estimator  vanishes logarithmically  with the  total
observation  time. In  other words,  the diffusion  constant estimated
from  one trajectory  by this  method tends  toward the  correct value
logarithmically slowly.   The prefactor $C(H)$, which  is displayed in
Fig.\ref{fig2},   reaches   a   minimum  at   $H^*\simeq0.30$.    From
Fig.\ref{fig2},  we notice  that,  keeping  the resolution  $\epsilon$
fixed, the  variance of $u$  will be small  for processes with  $H \in
[0.15,0.6]$,  typically.  This  interval  encompasses  almost all  the
anomalous    exponent    values    reported   in    single    particle
studies. Conversely, the function  $C(H)$ diverges as $H\rightarrow 0$
or $1$. Therefore, we can expect that, even with the ergodic choice of
$\alpha$, the estimates of the diffusion constant should become highly
inaccurate for nearly localized or nearly ballistic fBm processes.

In  conclusion,   we  have  shown  that   the  true,  ensemble-average
generalized diffusion coefficient $K$  of a fractional Brownian motion
of  known Hurst index $H$  can be obtained  from single trajectory
data using  the weighted least-squares  estimator in Eq.~(\ref{estim})
with the weight  function $\omega(t) = 1/(t_0 +  t)^{1+2 H}$.  Such an
estimator possesses an  ergodic property so that $K$  can be evaluated
with  any necessary  precision but  at the  expense of  increasing the
observation  time $T$  (or  decreasing $t_0$).   A  limitation of  the
present  class   of  estimators,   which  are  based   on  single-time
functionals of  $B_t^2$, is  admittedly their slow  convergence toward
the  ensemble  average.   Two-time  functionals,  based  on  the  time
averaged MSD,  for instance, exhibit faster convergence:  for fBm with
$H<3/4$ the  relative variance  of the time  averaged MSD  vanishes as
$t_0/T$ \cite{dengbarkai}.  Nevertheless  these other estimators might
be more sensitive  to measurement errors and may  not be accurate when
diffusion is no longer a pure  process but a mixture of processes with
different  characteristic times.   A  quantitative comparison  between
estimators  beyond the  ideal  cases considered  here  is a  necessary
future step.

\acknowledgments

GO acknowledges helpful discussions with M. Kleptsyna. DSD, CMM and GO
are  partially supported by  the ESF  Research Network  "Exploring the
Physics of Small  Devices". CMM is supported by  the European Research
Council and the Academy of Finland.

\end{document}